
\input phyzzx
\hsize=6.0in
\vsize=8.9in
\hoffset=0.0in
\voffset=0.0in

\FRONTPAGE
\line{\hfill BROWN-HET-830}
\line{\hfill March 1992}
\vskip.5truein
\titlestyle{SUPERCONDUCTING COSMIC STRINGS AND PRIMORDIAL\break
MAGNETIC FIELDS
}
\bigskip
\author{Robert H. Brandenberger$^{1)}$, Anne-Christine Davis$^{2)}$,
Andrew M. Matheson$^{1),3)}$ and Mark Trodden$^{4)}$}
\smallskip
\centerline{1) Physics Department, Brown University}
\centerline{Providence, RI  02912, U.S.A.}
\smallskip
\centerline{2) Department of Applied Mathematics
and}
\centerline{Theoretical Physics and Kings College}
\centerline{University of Cambridge}
\centerline{Cambridge CB3 9EW, U.K.}
\smallskip
\centerline{3) 2 Soldiers Field Park \#803}
\centerline{Harvard Business School}
\centerline{Boston, MA  02163, U.S.A.}
\smallskip
\centerline{4) Churchill College}
\centerline{University of Cambridge}
\centerline{Cambridge, U.K.}
\bigskip
\abstract

\noindent
We consider grand unified theories with superconducting cosmic strings
and which admit the mechanism for generating primordial magnetic
fields recently discussed by Vachaspati. We show that these models are
severely constrained by cosmological arguments. Quite generically,
either stable springs or vortons will form. Provided the mass per unit
length of the strings is sufficiently large, these stable
configurations will overclose the Universe.
\endpage
\chapter{Introduction}
The possibility [1]
that some models incorporating cosmic strings [2] may, under certain
conditions, act as superconducting loops and carry large currents
(large meaning currents of the order of magnitude of the string mass
scale, expressed in appropriate units), spawned a rash of speculation
on the possible manifestations of ``superconducting strings" (for a
review see e.g. Ref. 3).  Missing
from the menagerie of bizarre cosmological exotica, however, was a
mechanism to provide the requisite magnetic fields that would act as
seeds for the string currents.

Recent work [4] has illuminated one way of generating primordial
magnetic fields at very early times in the evolution of the universe.
Indeed, it is possible that coherent magnetic fields (\underbar{B}
fields hereafter) could be generated in sub-horizon sized patches
during a large class of grand unified transitions.  An obvious
consequence of producing these \underbar{B}-fields would be to charge
up any superconducting string loops, quite conceivably formed in the
same transition that gives rise to the \underbar{B}-fields.

It is the ultimate fate of these primordial loops of string that
concerns us in this note.  If no \underbar{B}-fields are present, the
loops will shrink and disappear.  If \underbar{B}-fields are present,
and have a coherent component that extends over a region as large as
the loop, the loop will develop a current as it collapses.  The
current will induce forces tending to resist the string tension of the
loop [5], hence diminishing the rate of decay of the loop and prolonging
its life. However, the current cannot exceed a maximal value $j_{max} \sim e
\sqrt{\mu}$ (where $\mu$ is the mass per unit length of the string),
since if $j > j_{max}$ it becomes energetically favourable for the charge
carriers to jump off the string, leading to strong electromagnetic radiation
[6]. There are hence two possible ultimate fates for a string loop. If
the current can increase to $j_{max}$, then the loop will decay by emitting
a burst of electromagnetic radiation. If, on the other hand, the current
$j_{spring} \sim \sqrt{\mu}$ [5] for which the force induced by
electromagnetism balances the tension is smaller than $j_{max}$, then stable
loops (``springs" [5]) result. The fate of the loop is a parameter-dependent
question.  In some models, stable springs will form, while in others they
will not [5, 7].

There is a second and more generic class of stable string configurations:
vortons [8 - 11]. These result from strings with a net charge and
induced angular momentum. As discussed in [11], strings with net
charge will be generically produced by intersections of current
carrying strings. The charge acts to further stabilize the configurations
against collapse. This provides a mechanism by which stable objects form
even if $j_{spring} > j_{max}$.

Our claim is that all cosmic string models that admit springs or
vortons, and
which lead to primordial \underbar{B}-fields at temperatures above a
critical temperature $T_i$ (to be determined below) are ruled out.
We can eliminate these theories because, as Copeland et al. observed
[5], stable springs produce an overdensity of matter in the same
manner as do primordial magnetic monopoles [12].  A necessary caveat
is that the mass per unit length of strings should be sufficiently
large (This bound will be discussed at the end of the Letter). If
the strings are to have some relevance for galaxy formation, their
mass per unit length $\mu$ must be about $10^{21} kg m^{-1} \,\bigl(
(10^{16} GeV)^2\bigr)$.  In many particle physics models the strings could
be much lighter, but such strings are, by and large, uninteresting
for cosmology, and we exclude them from consideration.

The outline of this Letter is as follows. We first review Vachaspati's
mechanism [4] for the generation of magnetic fields. Next, we study
the increase of currents on a cosmic string loop and determine the
mass density in cosmic springs at the time of
nucleosynthesis. Demanding that this density not exceed the energy
density in radiation leads to new cosmological bounds on particle
physics models.

\chapter{Primordial Magnetic Fields}

To establish our result, we recap the work of [4]. Under very general
conditions (that gauge fields acquiring mass during spontaneous
symmetry breaking carry electric charge), a magnetic field
\underbar{B},
smooth on a length scale $r_i$, is expected to form during a phase
transition at temperature $T_i$, with $B(T_i ) \sim T_i^2$ and $r_i
\sim {1\over T_i}$.  Using the temperature-time relationship
appropriate to the radiation-dominated era (during which we
take the transition to occur), this coherence length is a fraction
${\lambda T_i\over M_{pl}}$ of the horizon size $t_i (T_i)$.  As usual,
$M_{pl}$
is the Planck mass and $\lambda^2 \equiv {4\pi^3\over 45} g^*$ is a
constant determined by the effective number of spin degrees of freedom
$g^*$ of the theory under consideration.  Typically, $g^* \geq
200$
and $\lambda^2 \geq 500$.

Since the cosmological plasma is a very good conductor on the large scales we
are interested in, we expect the lines of magnetic flux to be frozen into the
plasma and the patches to retain a constant comoving size.
Then, the magnetic field within a coherence
length scales as
$$
B(t) = B(t_i) \left({{t_i} \over t} \right) .\eqno\eq
$$
Equivalently, the flux through a single patch in which the magnetic
field is uniform is conserved:
$$
\Phi_{patch} (t) = \pi r_i^2 B(t_i) .\eqno\eq
$$
On scales larger than the patch size, the flux has random orientation.
Hence, the net flux through a horizon size loop enclosing $N^2$
patches each carrying a flux $\Phi_{patch}$ will be of the order
$$
\Phi_{net} \sim N \Phi_{patch} \sim \left( {t \over {t_i}}
\right)^{1/2} {{t_i} \over {r_i}} \Phi_{patch} \eqno\eq
$$
(There are $(t_i / r_i)^2$ patches at time $t_i$ enclosed by a horizon
size loop).

\chapter{Springs}

The superconducting property of superconducting strings means that if
a loop encloses a flux $\Phi_0$ at the time $t_0$ when the loop forms, it will
generate a current to preserve the flux threading it at all later
times.  There are thus two contributions to the total flux threading a loop:
$\Phi_1$ which is the flux induced by the coherent magnetic fields and
$\Phi_2$ which is the flux induced by the current on the string. It is clear
that
$$
\Phi_2(t) = 2\pi R(t) J(t) \eqno\eq
$$
and so flux conservation:
$$
{d\over dt}[\Phi_1 + \Phi_2] = 0 \eqno\eq
$$
yields our basic differential equation
$$
{dJ\over dt} = \, - {J\over R} \, {dR\over dt} - {1\over 2\pi R} \,
{d\Phi_1\over dt}\eqno\eq
$$
where $J$ is the current, $R$ is the size of the loop, and
$\Phi_1$ is the flux through the loop (hereafter written as $\Phi$).
In (3.3), the first term on the right hand side represents current build-up
due to the collapse of the loop, while the second term represents current
build-up due to the time-variation of the primordial \underbar{B}-field.
{}From (3.2), we obtain
$$
{d\over dt} \, (RJ) = - {1\over 2\pi} \, {d\over dt} \, (\Phi
)\eqno\eq
$$
or
$$
J(t) = {1\over 2\pi R(t)} \, \bigl[ - \Phi (t) + {\rm
const}\bigr]\eqno\eq
$$
Now, as $J(t) = 0$ initially, and since $R (t_0 )$ is finite,
$$
J(t) = {1\over 2\pi R(t)} \, \bigl[ \Phi (t_0 ) - \Phi (t)
\bigr].\eqno\eq
$$

To interpret (3.6) we need to put in expressions for $R(t)$
(ignoring the back reaction of the current on $R(t)$) and $\Phi (t)$.
The evolution of
large loops is friction dominated immediately after the cosmic string
producing phase transition. At later times, the decay of $R(t)$
is due to gravitational radiation [13]:
$$
R(t) = R(t_0 ) - \gamma G {\mu} (t - t_0 )\eqno\eq
$$
with $\gamma \sim \, 100$.

Consider now the flux through a loop of initial radius $R(t_0)$ formed
at time $t_0$. From (2.3), the initial flux at time $t_0$ will be
$$
\Phi (t_0) = \Phi_{patch} {{R(t_0)} \over {r_i (t_0/t_i)^{1/2}}}
= \pi r_i B(t_i) R(t_0) \left({{t_i} \over {t_0}} \right)^{1/2} .
\eqno\eq
$$
At later times $t > t_0$, the number of coherence regions inside the
loop will drop and
$$
\Phi (t) = \Phi_{patch} {{R(t)} \over {r_i (t/t_i)^{1/2}}}
= \pi r_i B(t_i) R(t) \left({{t_i} \over {t}} \right)^{1/2} .
\eqno\eq
$$
Hence, from (3.6)
$$
J(t) = {1 \over 2} r_i B(t_i) {{R(t_0)} \over {R(t)}} \left({{t_i}
\over {t_0}} \right)^{1/2} \left[ 1 - {{R(t)} \over {R(t_0)}} \left(
{{t_0} \over t} \right)^{1/2} \right] \equiv J_0 {{R(t_0)} \over {R(t)}}
\left[ 1 - {{R(t)} \over {R(t_0)}} \left(
{{t_0} \over t} \right)^{1/2} \right] \eqno\eq
$$
with
$$
J_0 = {1 \over 2} r_i B(t_i) \left( {{t_i} \over {t_0}} \right)^{1/2} .
\eqno\eq
$$

The time dependence of $J(t)$ can be inferred from (3.10). Since
$R(t)$ decreases only very slowly, the current $J(t)$ rapidly rises to
the value $J_0$ and remains approximately constant until (see (3.7)) the
time
$$
t_d = (\gamma G \mu)^{-1} R(t_0) ,\eqno\eq
$$
at which time the current once again increases sharply during the
period when the loop radius decreases from the value
$\gamma G \mu R(t_0)$ to a final value at which the loop becomes a
spring, or when the maximal current is reached, whichever occurs
first. During this final collapse phase, the product $R(t) J(t)$ is
constant.

Using the values
$$
r_i = \alpha_1 T_i^{-1} \,\,\,\, {\rm and} \,\,\, B(t_i) = \alpha_2
T_i^2 \eqno\eq
$$
for the initial coherence length and magnitude of the magnetic field
(where $\alpha_1$ and $\alpha_2$ are constants which depend on the
details of the phase transition), then in the case
$t_0 =
t_1$ when the cosmic strings form in the same phase transition as the
magnetic field, the current $J_0$ becomes
$$
J_0 = {1 \over 2} \alpha_1 \alpha_2 T_i = \alpha_3 \sqrt{\mu} = x
j_{spring}, \eqno\eq
$$
where $x$ is a combination of $\alpha_1, \alpha_2$ and the constants
$j_{spring} / \sqrt{\mu}$ and $T_i / \sqrt{\mu}$. Thus, we conclude
that for $t_0 = t_1$ already the ``initial" current $J_0$ can be of
the right order of magnitude to give springs.

When $T_0 = T_2 < T_1$, where $T_2$ is the temperature of the cosmic
string producing phase transition, the relation between $J_0$ and the
spring current is unchanged since
$$
J_0 = {1 \over 2} \alpha_1 \alpha_2 T_2 = \alpha_3 \sqrt{\mu} = x
j_{spring}. \eqno\eq
$$

Consider now the case $x \ll 1$. In order for such a model with
$j_{spring} < j_{max}$ to be cosmologically safe, the current $J_0$
must be sufficiently small such that $J(t)$ remains smaller than
$j_{spring}$ until $R(t)$ shrinks to a size comparable to the
thickness $w \sim (\mu){-1/2}$ of the string, independent of the
temperature $T_0 < T_2$ when the loops are split off from the network
of infinite cosmic strings.
A loop formed at temperature $T_0$ has a typical size $R_i =
\alpha t_0 (T_0 )$, equal to
$$
R_i = {\alpha\over\lambda} \, {M_{pl}\over T_0^2}, \eqno\eq
$$
and develops a current
$$
J_0 = x {{T_0} \over {T_2}} j_{spring} \eqno\eq
$$
due to the magnetic flux threading the loop.
While collapsing to size $w$, the current increases by a factor
$$
f = {{R_i} \over w} = {{\alpha M_{pl}} \over {\lambda T_0^2 w}} .\eqno\eq
$$
For the current at this point not to exceed $j_{spring}$ we must have
$$
x {{T_0} \over {T_2}} f = x {\alpha \over \lambda} {{M_{pl} w^{-1}}
\over {T_2 T_0}} < 1 .\eqno\eq
$$

At high temperatures
immediately after the cosmic string producing phase transition, the
motion of strings is dominated by the friction of the surrounding
plasma. In the absence of primordial \underbar{B}-fields,
the loops formed would simply shrink and disappear.
However, in the presence of primordial \underbar{B}-fields these loops
can only shrink to a final radius
$$
R_f = x {\alpha\over\lambda}  {M_{pl}\over T_0} {1 \over T_2} \eqno\eq
$$
when $J = j_{spring}$, and then become springs.
Our results
therefore apply equally well to the friction-dominated case.

{}From (3.19), we see that
independent of how small the constants $x$ and $\alpha / \lambda$ are
(according to the recent cosmic string evolution simulations [14]
$\alpha \sim 10^{-3}$),
eventually $T_0$ will become sufficiently small such that the
condition is violated, and hence springs will form. The presence of
springs causes cosmological problems. They might overclose the
Universe (see e.g. Ref. [15]). More specifically, successful
nucleosynthesis requires that
Universe is still radiation dominated at the time of nucleosynthesis.
The only ways to satisfy this constraint are to make the strings
sufficiently light, or to make sure that no \underbar{B}-fields are
produced above a certain temperature.

Taking $T_0$ to be $T_2$ or the temperature when the inequality (3.19)
is marginal (whichever is lower), then
the mass of the earliest spring, whose radius is given by (3.20), is
$$
M_{loop} = 2 \pi \mu R_f \sim 2 \pi {\alpha\over\lambda} x {M_{pl}\over T_0}
\sigma \eqno\eq
$$
where $\sigma = \sqrt{\mu}$ is the scale of symmetry breaking.
If there is one loop of radius $R_i$ per horizon volume
at temperature $T_0$ which forms a spring (it is such loops which
dominate the total mass - see e.g. Ref. [16]), then
there will be $\bigl( {T_0\over T_F}\bigr)^3$ springs per horizon volume
at a lower temperature $T_F$, and a total mass of springs inside the
horizon of
$$
M_{tot} = 2 \pi {\alpha\over\lambda} \,x\, {M_{pl}\over T_0} \sigma \bigl(
{T_0\over T_F}\bigr)^3\eqno\eq
$$
In particular, if the Universe is to be radiation-dominated at
nucleosynthesis when $T_F = 0.1 MeV$, the energy in springs must be
less than that in radiation.  We therefore require that the ratio
$$
4 \pi \alpha x
{\sigma\over T_F} \Bigl( {T_0\over M_{pl}}\bigr)^2 << 1 .\eqno\eq
$$

If $x \approx 1$ and $T_0 = T_2$, radiation-dominated
nucleosynthesis implies that $T_2 << 10^{11} GeV$ if \underbar{B}-
fields are present. For
strings with potential to form galaxies, the temperature at which the
strings are formed is
$T_2 \approx 10^{15} GeV$.
Our constraint (3.23) (this time evaluated with $T_0 \neq T_2$)
therefore implies no
\underbar{B}-fields may be produced in superconducting spring theories
above $T_i \equiv T_0 \sim 10^{10} GeV$.  For radiation-domination down
to $T=1eV$, our limit is $T_0 \leq 3 \times 10^7 GeV$. If, on the
other hand, we assume $T_0 = \sigma \neq 10^{15}GeV$, then (3.23)
yields the constraint $T_0 < 10^{12}GeV$ for radiation domination to
$T_F = 0.1MeV$.

\chapter{Vortons}

Let us now consider vortons, a class of stable field configurations
more generic than springs, arising also if $j_{spring} > j_{max}$.
The essential difference between vortons
and the springs considered above is that the vortons have charge and angular
momentum [8 -10].
This feature leads to the vorton solutions being intrinsically more stable than
the corresponding spring solutions and therefore posing a more serious
cosmological problem in the type of theories considered above.

We consider
as a simple example the bosonic $U(1) \times U(1)_{em}$ model originally
proposed by Witten [1]. The Lagrangian contains two scalar fields,
${\hat \phi}$ which gives rise to the strings and
${\hat \sigma}$ which condenses inside the string and induces
the superconducting current.

Defining $\phi\equiv|{\hat \phi}|$ , $\sigma\equiv|{\hat \sigma}|$ , and
$\theta\equiv$arg${\hat \sigma}$, there exist two currents on the vortex
worldsheet, one which is conserved topologically and gives the string a
current
$$
{\tilde j_{a}} = \epsilon_{ab}\partial^{b}\theta, \eqno\eq
$$
and a second which is conserved by the equations of motion and gives
the vortex an electric charge
$$
j_{a} = \sigma^{2}(\partial_{a}\theta - A_{a}) \eqno\eq
$$
($A_a$ being the electromagnetic potential) with respective charges
$$
N = 2\pi R n \,\,\,\,\,\,\,\,\,\, Q = \oint dl\int_{X}\sigma^{2}(\omega -
A_{0})
.\eqno\eq
$$
Here, $Q$ is the electromagnetic charge, and $R$ is the average radius of
the loop. Also $\omega\equiv\partial_{0}\theta$ and $\int_{X}$ is the
integral over the string cross-section.

The presence of a non-zero $Q$ on the string causes the loop to have angular
momentum, and in the case of {\it chiral} vortons
for which $\omega = n$, we have
$$
\theta = \omega (t-l),\eqno\eq
$$
where $l$ is the arc length around the loop. Since $\omega/n =1$ can be shown
[8] to be an attractor, chiral vortons are expected to dominate, and hence
we shall focus on them.

For chiral vortons, the total electromagnetic charge on a loop is proportional
to the topological charge $N$. It will prevent the loop from collapsing and
there will be a finite stabilization radius $R_F$ which can be computed by
minimizing the total static energy [10] given by:
$$
E = 2\pi R\mu + 4\pi {N^{2} \over R} \Sigma \eqno\eq
$$
where $\Sigma$ is an integral of $\sigma^{2}$ over the string cross section, to
yield
$$
R_F \simeq {N \over {\sqrt{4 \pi} \sigma}} .\eqno\eq
$$

Models with vortons can cause cosmological problems if $R_F$ is larger
than the width $w \sim \sigma^{-1}$ of the string. In the other case,
the loop will simply disappear in a burst of elementary quanta. A loop of
initial radius $R_0$ will carry an average initial topological charge
$N$ of
$$
N = {{R_0} \over {\xi(T_0)}}, \eqno\eq
$$
where $\xi(T_0) \sim T_0^{-1}$ is the correlation length of the phase
of $\hat{\phi}$ at the time of loop formation.
Hence,
$$
R_F = {{\alpha} \over {\sqrt{4 \pi} \lambda}} {{M_{pl}} \over {T_0}}
{1 \over \sigma} \eqno\eq
$$
which, up to factors of order unity, agrees with the final radius (3.20)
of cosmic springs.

Equation (4.8) leads to interesting consequences. If $R_F < {1 \over
\sigma}$, then no stable vortons can form. With the values of $\alpha$
and $\lambda$ given before, this occurs for $T_0 > 10^{14} GeV$. In
models with $\sigma > 10^{14} GeV$, vortons will hence not immediately
form, but only once the temperature of the Universe drops below this
value. If $\sigma < 10^{14} GeV$, vortons form immediately after the
phase transition.

Since the stabilization radius of a vorton agrees with the final
spring radius, the mass bounds are identicial to those discussed at
the end of Section 3. We conclude that all theories with $\sigma >
10^{10} GeV$ admitting superconducting cosmic strings are ruled out;
those with $j_{max} > j_{spring}$ by springs, those with $j_{max} <
j_{spring}$ by vortons.

In addition, the presence of charge on strings modifies the
considerations of Section 3. There are now two contributions to the
total current, that coming from the string current which we denote by
$J_c$ and that coming from the charge (denoted by $J_Q$). The final
spring radius $R_f$ is now determined by equality between total
current and spring current. Hence, for given initial winding number
and radius, $R_F$ is larger with charge than without. Hence, the mass
bounds are stronger.

First, we determine for which $J_Q$ substantial changes to the results
of Section 3 can be expected. This will be the case if for loops of
radius $R_F$ (see (3.20)) the current $J_Q$ exceeds $j_{spring} \sim
\sigma$. With [10]
$$
Q \sim e {{R_i} \over {\xi(T_0)}} \eqno\eq
$$
and
$$
J_Q(t) \sim {{\omega Q} \over {2 \pi}} \eqno\eq
$$
we see that for
$$
T_0 < {e \over {4 \pi^2}} {\alpha \over \lambda} M_{pl} \sim 10^{13}
GeV \eqno\eq
$$
there will be a substantial modification of the previous results. The
new stabilization radius $\tilde{R_F}$ is determined by $J_Q = \sigma$
which gives
$$
\tilde{R_F} = {e \over {4 \pi^2}} \bigl({\alpha \over \lambda}\bigr)^2
\bigl({{M_{pl}} \over {T_0}}\bigr)^2 {1 \over \sigma} . \eqno\eq
$$

To determine the new constraints which result when taking into
account vortons, we redo the mass estimates of (3.21 - 3.23) with
$\tilde{R_F}$ instead of $R_F$. A simple calculation gives the
following results: for $\sigma = 10^{15} GeV$, the temperature $T_0$
at which the magnetic field is created is bounded by
$$
T_0 < 2 \times 10^6 GeV \eqno\eq
$$
(with $T_F = 0.1 MeV$). Alternatively, if we assume that $T_0 = \sigma
\neq 10^{15} GeV$, the constraint becomes
$$
T_0 < 3 \times 10^{10} GeV. \eqno\eq
$$
These bounds are several orders of magnitude stronger than the ones
obtained when neglecting vortons.

\chapter{Conclusions}

To summarize, we have derived constraints on models admitting
superconducting cosmic strings. We have assumed that at some
temperature $T_0$, primordial magnetic fields are produced [4] during
one of the phase transitions which breaks the original symmetry group
down to the standard model. The constraints come from demanding that
stable springs or vortons do not cause the Universe to become matter
dominated before nucleosynthesis.

Focusing on cosmic string models with a scale of symmetry breaking
$\sigma \sim 10^{15}GeV$ (the value required in galaxy formation
scenarios [17]), we distinguish the following cases. If the spring
current $j_{spring}$ is larger than the maximal current $j_{max}$,
then there are only stable vortons and the constraint on $T_0$ is
$T_0 < 10^{10}GeV$. For $j_{spring} < j_{max}$, both stable springs
and vortons exist. Vortons dominate the mass density if $T_0 <
10^{13}GeV$, springs are more important otherwise. In such models, the
Universe is matter dominated at nucleosynthesis unless $T_0 < 2 \times
10^6GeV$.

For cosmic strings
of relevance for structure formation, the phase transition producing
\underbar{B}-fields cannot occur much above the electroweak scale. If
such fields are produced at the grand unified scale, then the mass per
unit length of the superconducting cosmic strings cannot be
sufficiently large for the strings to be of relevance in structure
formation scenarios.

In analyzing models with $j_{spring} > j_{max}$, we have assumed that
the charge on a loop remains constant while it is emitting
electromagnetic radiation. In a perturbative analysis, this is
obvious. However, nonperturbative processes may allow net charge
decrease (e.g. by further loop fragmentation).
\bigskip
\centerline{\bf Acknowledgements}
\medskip
This work has been supported in part by DOE grant DE-AC02-76ER03130
Tasks A \& K, by an Alfred P. Sloan Fellowship to R.B., and by an NSF-
SERC Collaborative Research Award NSF INT-9022895 and SERC GR/G37149.
\endpage

\centerline{\bf REFERENCES}

\item{1.}
E. Witten, {\it Nuc. Phys.} {\bf B249} (1985) 557.
\item{2.}
T. W. B. Kibble, {\it J. Phys} {\bf A9} 1387; \nextline
A. Vilenkin, {\it Phys. Rep.} {\bf 121} 263.
\item{3.}
E. Copeland, {\it Cosmic Strings and Superconducting Cosmic Strings},
in Proc. of Second Erice Summer School on Dark Matter, Erice, May 1988
(World Scientific, Singapore, 1988).
\item{4.}
T. Vachaspati, {\it Phys. Lett.} {\bf 265B} (1991) 258.
\item{5.}
E. Copeland, M. Hindmarsh and N. Turok, {\it Phys. Rev. Lett.} {\bf 58}
(1987) 1910.
\item{6.}
G. Field and A. Vilenkin, {\it Nature} {\bf 326} (1986) 944;\nextline
J. Ostriker, C. Thompson and E. Witten, {\it Phys. Lett.} {\bf 180B}
(1986) 231;\nextline
D. Spergel, T. Piran and J. Goodman, {\it Nucl. Phys.} {\bf B291}
(1987) 847;\nextline
E. Copeland, D. Haws, M. Hindmarsh and N. Turok, {\it Nucl. Phys.}
{\bf B306} (1988) 908.
\item{7.}
A. Babul, T. Piran and D. Spergel, {\it Phys. Lett.} {\bf 202B} (1988)
307;\nextline
C. Hill, H. Hodges and M. Turner, {\it Phys. Rev.} {\bf D37} (1988)
263;\nextline
D. Haws, M. Hindmarsh and N. Turok, {\it Phys. Lett.} {\bf 209B}
(1988) 255.
\item{8.}
R. Davis and E.P.S. Shellard, {\it Phys. Lett.} {\bf 207B} (1988)
404;\nextline
R. Davis and E.P.S. Shellard, {\it Phys. Lett.} {\bf 209B} (1988) 485.
\item{9.}
R. Davis, {\it Phys. Rev.} {\bf D38} (1988) 3722.
\item{10.}
R. Davis and E.P.S. Shellard, {\it Nucl. Phys.} {\bf B323} (1989) 209.
\item{11.}
T.J. Allen, {\it Phys. Lett.} {\bf 231B} (1989) 429;\nextline
T.J. Allen, {\it Phys. Lett.} {\bf 250B} (1990) 29.
\item{12.}
M. Khlopov and Ya. B. Zel'dovich, {\it Phys. Lett.} {\bf 79B} (1978)
239;\nextline
J. Preskill, {\it Phys. Rev. Lett.} {\bf 43} (1979) 1365.
\item{13.}
N. Turok, {\it Nucl. Phys.} {\bf B242} (1984) 520;\nextline
T. Vachaspati and A. Vilenkin, {\it Phys. Rev.} {\bf D31} (1985)
3052;\nextline
C. Burden, {\it Phys. Lett.} {\bf 164B} (1985) 277.
\item{14.}
A. Albrecht and N. Turok, {\it Phys. Rev.} {\bf D40} (1989)
973;\nextline
D. Bennett and F. Bouchet, {\it Phys. Rev. Lett.} {\bf 60} (1988)
257;\nextline
B. Allen and E.P.S. Shellard, {\it Phys. Rev. Lett.} {\bf 64} (1990)
119.
\item{15.}
A.-C. Davis and R. Brandenberger, {\it Cosmological Limits on Stable
Particle Production at High Energy}, Brown preprint BROWN-HET-828
(1991) Phys. Lett. B (in press).
\item{16.}
T. Vachaspati and A. Vilenkin, {\it Phys. Rev.} {\bf D30} (1984)
2036;\nextline
D. Mitchell and N. Turok, {\it Phys. Rev. Lett.} {\bf 58} (1987) 1577.
\item{17.}
N. Turok and R. Brandenberger, {\it Phys. Rev.} {\bf D33} (1986)
2175;\nextline
A. Stebbins, {\it Ap. J. (Lett.)} {\bf 303} (1986) L21;\nextline
H. Sato, {\it Prog. Theor. Phys.} {\bf 75} (1986) 1342.
\end